# Networks of Border Zones: Multiplex Relations of Power, Religion and Economy in South-Eastern Europe, 1250-1453 AD

**Johannes Preiser-Kapeller**
Austrian Academy of Sciences, Austria. *Johannes.Preiser-Kapeller@oeaw.ac.at*

**Abstract:**
*We demonstrate the application of the "multiplex networks"-approach for the analysis of various networks which connected individuals and communities in the politically highly fragmented late medieval Balkans (1204-1453 AD) within and across border zones. We present how we obtain relational data from our sources and the integration of these data into three different networks (of roads, state administration and ecclesiastical administration) of various topologies; then we calculate several indicators for influences and overlaps between these different networks which connect the same set of nodes (settlements). We analyse changes and continuities in the topologies of the various networks for three time-steps (1210, 1324 and 1380 AD) and demonstrate the role of these networks as frameworks for social interactions. Finally, we combine all three networks into one network which shows properties observed for the "small world" model. Thus, we demonstrate possibilities for capturing historical complexity with the help of the "multiplex networks" approach.*

**Note:** Colour versions of all figures are available in the online-version of this paper

## Fragmentation, Borderlands and the Concept of Multiplex Networks

The centuries after the fall of Constantinople to the Crusaders in 1204 were characterized by the political fragmentation of the former imperial sphere of Byzantium; attempts to establish hegemony by one of the local powers were followed by phases of disintegration of these polities until the Ottomans restored "imperial unity" (Laiou 2006). While political border zones frequently changed, religious denominations tried to preserve or expand their spheres of influence within the entire Balkans; furthermore, local and regional trading networks criss-crossed the region and integrated it in the late medieval "World system" (Abu-Lughod 1989). Therefore, political, religious and economic spheres of influence of the various centres of power were not congruent, but influenced each other.

The concepts of network analysis allow us to understand relations between different communities and authorities in a novel way; recent research has made clear, that various fields of relationships between individuals, communities and polities also span different networks. Munson and Macri examined different kinds of relationships between Maya centers and analyzed the "intersection of antagonistic, diplomatic, subordinate, and kinship" networks (Munson and Macri 2009). Maoz in his analysis of the "Evolution of International Networks" modelled networks of alliance, trade, common membership in international organisations and enmity between states (Maoz, 2011).

One of the technically most sophisticated analyses of multiplex networks between individuals so far was executed by Michael Szell, Renaud Lambiotte and Stefan Thurner from the Vienna Complex Systems Research Group; they state:

*"Human societies can be regarded as large numbers of locally interacting agents, connected by a broad range of social and economic relationships. (…) Each type of relation spans a social network of its own. A systemic understanding of a whole society can only be achieved by understanding these individual networks and how they influence and co-construct each other (…) A society is therefore characterized by the superposition of its constitutive socio-economic networks, all defined on the same set of nodes. This superposition is usually called multiplex, multi-relational or multivariate network."* (Szell et al. 2010; Szell and Thurner 2010; on multiplex networks cf. also Wasserman and Faust 1994, 73-75, 81-83, 145-146, 730; Berlingero et. al. 2011; Jackson 2008, 21, 47, 84; Newman 2010, 110-111).

The data for the study of Szell et al. (2010) stems from a massive multiplayer online-game. During a period of three years, almost every interaction of more than 300,000 players was registered; this analysis of a virtual society allows not only for an unprecedented abundance of material, but also for an unambiguity of data, since players could mark other individuals as friends or enemies, for instance (Szell et al. 2010).

**Quantity and Quality of Data: Networks of Roads and Administration**

A historian or an archaeologist will never have access to data of a quantity or quality comparable with that of Szell et al. (Brughmans, forthcoming). But for our case study, we tried to define a set of nodes in a combination with pieces of evidence which would allow us to span different networks within the same sample in relative consistency. Fortunately, there exist sources which cover a specific set of nodes with information on almost all ties relevant for a peculiar kind of relationship: these are documents from state and ecclesiastical administration on the one hand and descriptions of routes and roads on the other hand (Asdracha 1976; Darrouzès 1981); both categories of sources span networks across a set of localities. For our study, we chose one of the remaining core-regions of the Late Byzantine Empire, the area of Thrace, for which there also exist two volumes from the series "Tabula Imperii Byzantini" (written by Peter Soustal and Andreas Külzer), which aim at a comprehensive coverage of all aspects of geography and administration for all localities within a specific area (Külzer 2008; Soustal 1991; cf. also Popović forthcoming).

Soustal and Külzer had already reconstructed the networks of land and sea routes for these provinces for the late Byzantine period (Külzer 2008, 192–211; Soustal 1991, 132–148), which we adapted for our network analytical software (de Nooy et al. 2005): the street-network (Fig. 1) includes 119 nodes (or localities) and 350 links (or routes between nodes, whereupon we have included all routes without differentiating their relative importance for trans-regional or regional traffic; for (also more sophisticated) methods for the analysis of road networks Conolly and Lake 2006, 234–262; Gorenflo and Bell 1991, 80–98; Gregory and Ell 2007, 26–27; Popović in press), the sea route-network (Fig. 2), which of course only connects coastal cities, includes 32 nodes and 70 links (we made all links binary or reciprocal, since a route always leads in two directions). On this set of node we now intended to span the networks of administration of the state (Fig. 3) and of the church (Fig. 4) for a specific timeframe; on the basis of our sources, we chose the time around the year 1324 (Külzer 2008; Preiser-Kapeller 2008a; Soustal 1991). Both administrative networks include 119 nodes and 234 links; in contrast to the street-network, both networks decompose in two components, a larger one, which represents the sphere of influence of the Byzantine Empire respectively the Patriarchate of Constantinople, and a smaller one of the Bulgarian State respectively Patriarchate (Figs 3 and 4).

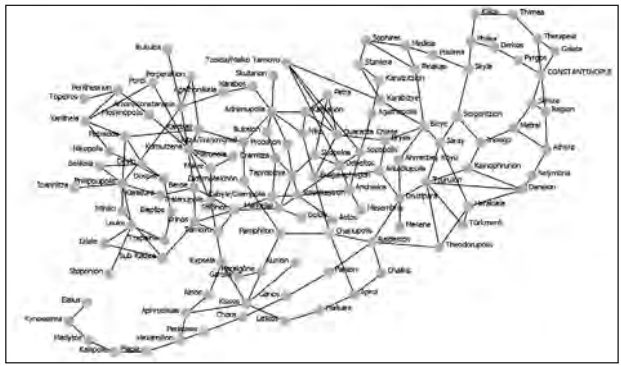

*Figure 1. Topological representation of the network of roads in the area of Thrace in the 13th/14th cent. (119 nodes or localities and 350 links or routes).*

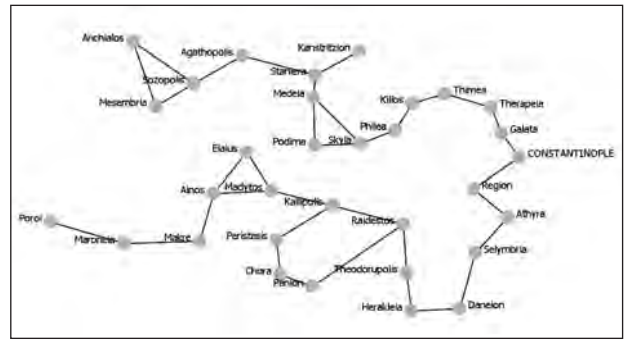

*Figure 2. Topological representation of the searoute-network for the area of Thrace in the 13th/14th cent. (32 nodes or localities and 70 links or routes).*

**Networks with Different Topologies**

We first analysed the different networks with regard to the standard measures on the level of individual nodes and of the entire network; then we compared the networks with regard to these measures and their overall topology. We compared the following measures:

- *average distance* measures the average path length between two nodes within the network (direct neighbours have a distance of 1).

- *clustering coefficient* measures the probability that two nodes connected with a node are connected to each other as well.

- *density* is the proportion of possible links that are actually present in a network (a network in which every node is connected with each other node has a density of 1).

- the *degree* of a node is the number of links connected to it; *degree centrality* indicates the relative significance of a node within a network due to its number of connections to other nodes.

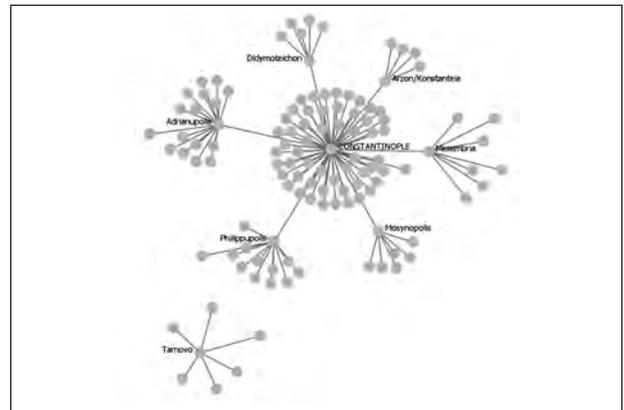

*Figure 3. Topological representation of the network of administration of the states for the area of Thrace in 1324 (119 nodes and 234 links between administrative centres and their subordinate localities).*

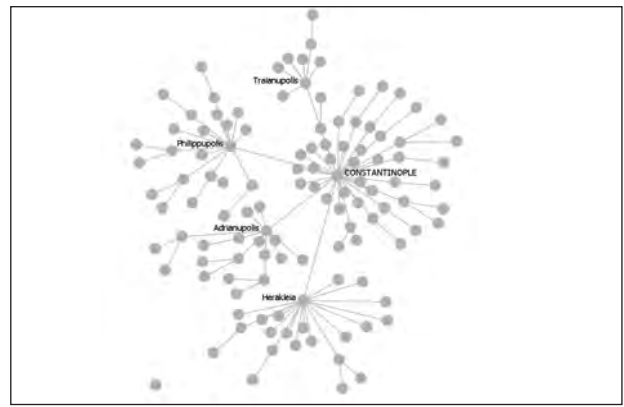

*Figure 4. Topological representation of the network of administration of the churches for the area of Thrace in 1324 (119 node and 234 links between the Patriarchate, metropolitan sees, archbishoprics and suffragan bishoprics).*

| Measure | Admin 1324 | Church 1324 | Street |
|---|---|---|---|
| Average Distance | 2.779 | 3.532 | 7.085 |
| Clustering Coefficient, Watts-Strogatz | 0 | 0 | 0,092 |
| Density | 0.017 | 0.017 | 0.025 |
| Link Count | 234 | 234 | 350 |
| Network Centralization, Betweenness | 0.836 | 0.865 | 0.196 |
| Network Centralization, Total Degree | 0.509 | 0.276 | 0.035 |

*Figure 5. A comparison of network analytical measures for the road, state and church networks for Thrace for 1324.*

- *betweenness centrality* measures the extent to which a node lies on paths between other nodes and indicates the relative significance of a node as "intermediary" within a network due to its position on many (or few) possible routes between other nodes.

- *centralization* "quantifies the range or variability" of the individual node indices within a network; in a network with a degree centralization of 1, all nodes are only connected to one central node, in a network with a betweenness centralization of 1, all nodes are equally connected through one central node (the network graph has the form of a star). (Newman 2010, 133–136, 168–169, 185–193, 199–204; Wasserman and Faust 1994, 101–103, 107, 171, 178–183, 188–192).

As the visualisation alone illustrates, these overlapping networks differ in many respects: the road-network is not only fully connected, but also shows a more decentralised, distributed architecture; the administrative networks are highly centralised and hierarchical, with links always leading from an administrative centre to its subordinates. These different topologies are also reflected in the most important network measures (Albert and Barabási 2002, 48–97; cf. also Jackson 2008, 56-65) with regard to centralisation in degree and betweenness, which are much higher for the administrative networks (Fig. 5); the higher centralisation of links also allows for a lower average distance between nodes. At the same time, the state and church networks are less densely spanned than the road-network; we do not observe any clustering in the star-like structures of the administrative ties, while we have a modest clustering coefficient in the road-network.

Such significant differences become also visible with regard to the frequency distribution of degree (Fig. 6 a-c): while in the road-network, the degree values are very equally distributed around their average value, the degree-distributions for the hierarchical administrative networks are much more unequal. The different architectures of these infrastructural networks can be very well connected to their different restraints, development and purposes: the network of roads and routes depends on the geographic environment, within which it shall provide connections among localities in a reasonable manner, while administrative hierarchies (of course depending on spatial factors as well) shall concentrate the flow of resources, information as well as loyalty and allegiance onto the centres of power and decision-making.

### Influences and Overlaps between Networks

The multiplexity of these "infrastructure" networks constructs a framework for political,

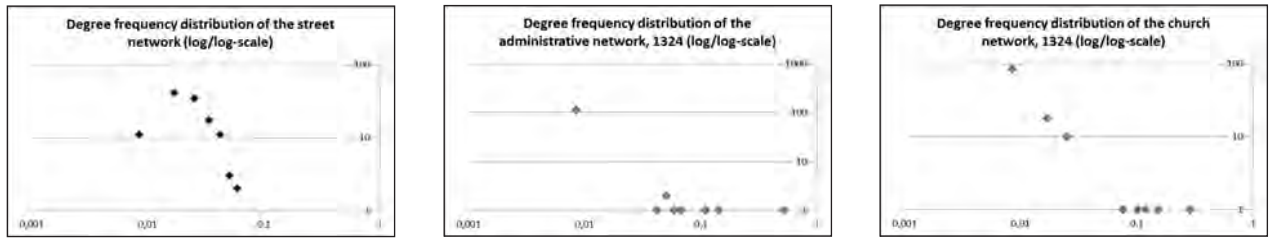

*Figure 6 a-c. A comparison of the road, state and church networks for Thrace for 1324: degree frequency distributions (on double-logarithmic scales; we measure the frequency of each degree values); on "degree", see section 3.*

economic, ecclesiastical etc. interaction; in order to analyse this framework further, we have to identify possible influences and overlaps between these networks (Fig. 7); we asked for instance to what amount the various networks overlap each other, more precisely: if two nodes, which are directly connected in one network, are also connected in one of the other or in all other networks (Maoz 2011, 39–41, 333–364; Munson and Macri 2009, 435–436; Szell et al. 2010; cf. also Berlingero et al. 2011). We determined the percentage of overlap of links between the three networks with the same number of nodes (streets, state, church) by means of a comparison of the matrices of the networks (Fig. 8): the highest overlap we detected, as could be expected because of their similar architecture, between the state and the ecclesiastical administrative networks for 1324; still, although they have the same number of links, they have only 15 % of all their links in common. But while the **overlap between state administration and street network is significantly lower, the ecclesiastical administrative networks and the street networks have almost the same percentage of links in common as the two administrative networks** – maybe due to the fact that the ecclesiastical framework still followed the Roman administrative organization from late antiquity, which in turn was connected to the Roman road system to a stronger degree than the administration of 14[th] century Byzantium, which had been adapted to more recent socio-economic and political changes with regard to the relevance of localities. Very low is in contrast the overlap between all three networks (Fig. 8). All in all, we see that even networks of very similar structure could link the same set of nodes in a fairly diversified manner.

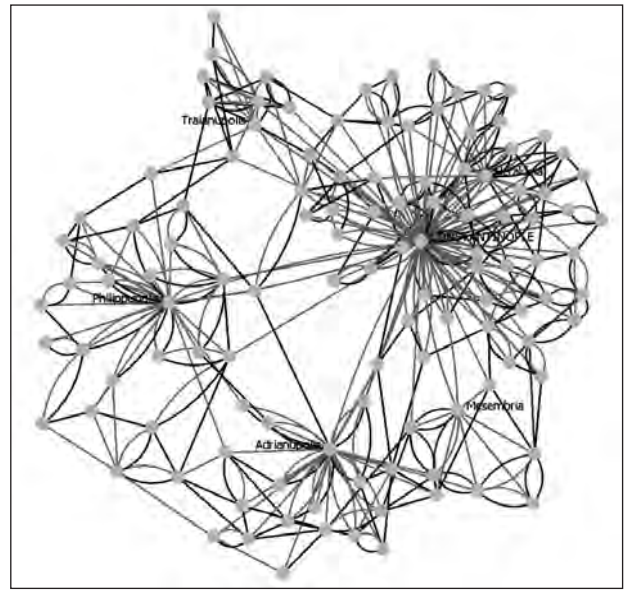

*Figure 7. The multiplex "infrastructure" network of Thrace, 1324 (for full colour image please see the online version of this paper where: black: street network; blue: searoutes-network; red: state administration network; green: church administration network).*

Another important question is if there exists a correlation between node degrees in different networks, or more explicit: do nodes which have many (few) links in a network have many (few) links in another network (Berlingero et al. 2011; Szell et al. 2010)? Again, we detected the highest degree correlation between the administrative networks (Fig. 9 a-b); significantly lower are the degree correlations between the administrative

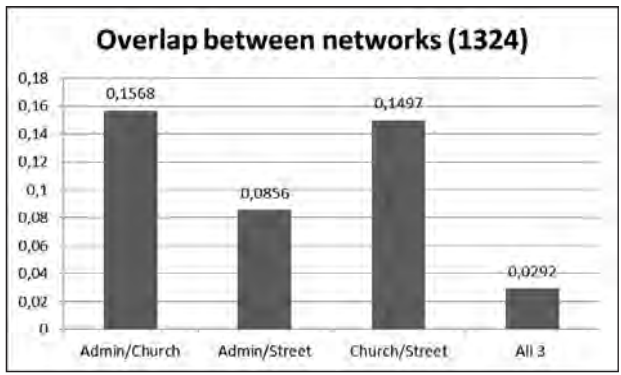

*Figure 8. The overlap (percentage of links between nodes common to two or all three networks) between the networks for Thrace in 1324.*

|  | Degree correlation (r/r²) 1324 | Between. correl. (r/r²) 1324 |
|---|---|---|
| **Admin/Church** | 0.8528/0.7273 | 0.9108/0.8295 |
| **Admin/Street** | 0.2384/0.0568 | 0.0678/0.0046 |
| **Church/Street** | 0.3042/0.0925 | 0.0825/0.0068 |

9a

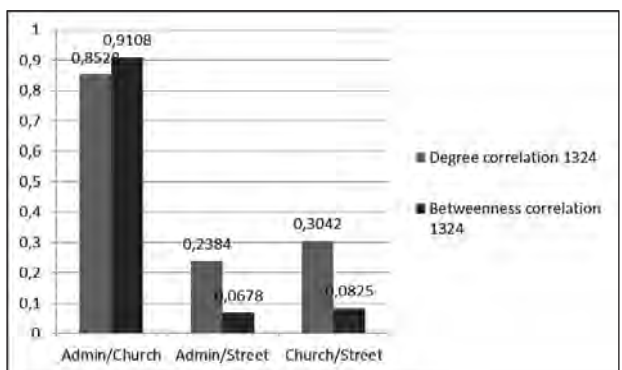

9b

*Figure 9 a and b. Degree and betweenness correlations between the networks for Thrace in 1324; on "degree" and "betweenness", see section 3.*

networks and the street networks (for a possible explanation of this phenomenon, see above our considerations on the overlaps between networks; this phenomenon may also be partly connected to the fact that the administrative hierarchies favoured specific localities, while we did not differentiate between routes of and nodes on regional or trans-regional relevance in our model of the street network). In a similar way, a determination of the correlations of values for the betweenness of nodes in various networks helps to indicate if a node which has an important intermediary position in one network has also a comparable position in another network; the results are similar for the correlation between the administrative networks, while the correlation between the administrative networks and the street network is even much more weaker, as could be expected because of the highly different betweenness-centralization of the respective networks (Fig. 9 a-b).

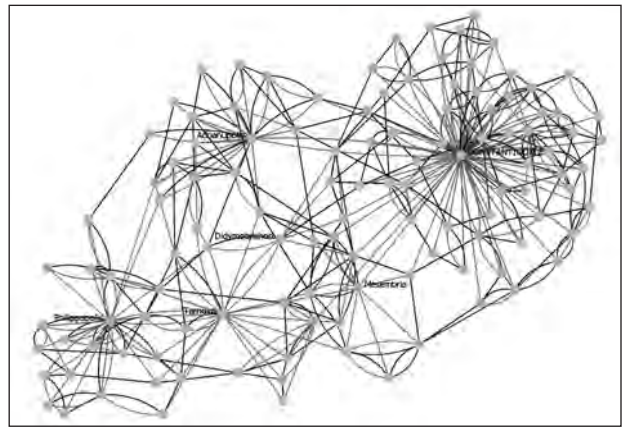

*Figure 10. The multiplex „infrastructure" network of Thrace, 1210 (for full colour image please see the online version of this paper where: black: street network; blue: searoutes-network; red: state administration network; green: church administration network).*

**Changes and Continuities of Networks in Time**

While in 1324, almost the entire territory was under the political and ecclesiastical control of Constantinople, we also constructed the administrative and ecclesiastical networks for two time points when this was not the case: in 1210, six years after the conquest of Constantinople, the Bulgarian Empire had been able to occupy large areas of Thrace, which also were subordinated to the Patriarchate in Tarnovo, while the newly established Latin Empire could only retain the regions around Constantinople, where bishops under the Latin

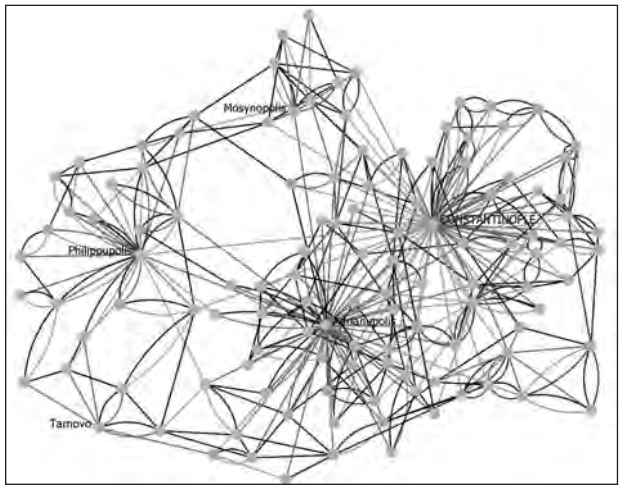

*Figure 11. The multiplex „infrastructure" network of Thrace, 1380 (for full colour image please see the online version of this paper).*

Patriarch were enthroned; so we encounter two large components in the administrative networks (Fig. 10) (Külzer 2008; Soustal 1991; Wolff 1948, 33–60).

Around 1380, the Ottoman Empire had already established itself as a strong power in Europe; from his new capital in Adrianople, Sultan Murad I controlled the majority of the Thracian cities, while Byzantium only retained the regions near their capital and the Bulgarian state some territories in the north (Külzer 2008; Preiser-Kapeller 2008a; Soustal 1991). So we identify three components in the state administrative network – but not in the ecclesiastical network, since the Patriarchate of Constantinople attempted to preserve its organizational framework under Ottoman rule (Fig. 11).

If we now compare central measures for the networks for the different time steps (Fig. 12 a-b), we detect a high degree of stability with regard to most values; for the historian, this is not a big surprise, since both Latin respectively Bulgarian and Ottoman invaders made much use of the existing organizational frameworks from Byzantine times. Yet, the political fragmentation in 1210 and 1380 and

| Measure | Admin 1210 | Admin 1324 | Admin 1380 |
|---|---|---|---|
| Average Distance | 2.543 | 2.779 | 2.528 |
| Clustering Coefficient, Watts-Strogatz | 0 | 0 | 0 |
| Density | 0.017 | 0.017 | 0.017 |
| Link Count | 234 | 234 | 232 |
| Network Centralization, Betweenness | 0.347 | 0.836 | 0.381 |
| Network Centralization, Total Degree | 0.466 | 0.509 | 0.397 |

| Measure | Church 1210 | Church 1324 | Church 1380 |
|---|---|---|---|
| Average Distance | 3.011 | 3.532 | 3.375 |
| Clustering Coefficient. Watts-Strogatz | 0 | 0 | 0 |
| Density | 0.014 | 0.017 | 0.017 |
| Link Count | 202 | 234 | 234 |
| Network Centralization. Betweenness | 0.192 | 0.865 | 0.899 |
| Network Centralization. Total Degree | 0.218 | 0.276 | 0.362 |

*Figure 12 a-b. A comparison of network analytical measures for the administrative networks in time, 1210 – 1324 – 1380.*

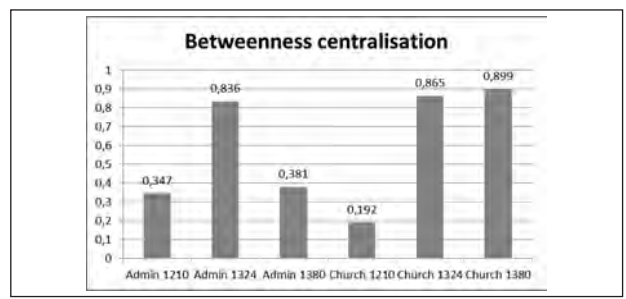

*Figure 13. A comparison of betweenness centralization for the administrative networks in time, 1210 – 1324 – 1380.*

the ecclesiastical fragmentation in 1210 are reflected in the significantly lower betweenness-centralization of administrative networks for these periods (Fig. 13).

Continuity we also observe for the degree

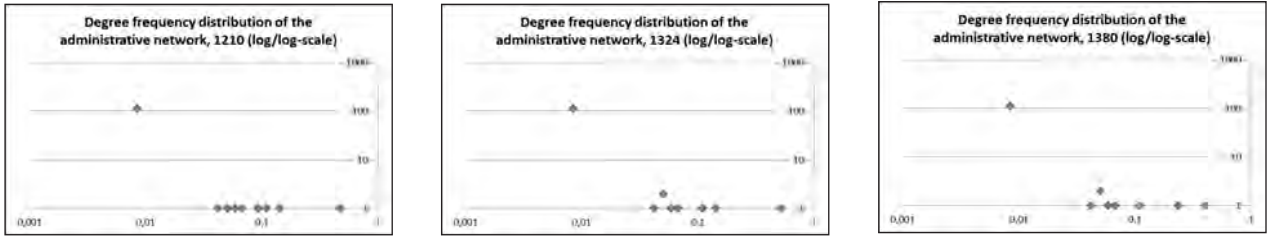

*Figure 14 a-c. The degree frequency distribution of the state administration network in time, 1210 – 1324 – 1380 (on double-logarithmic scales).*

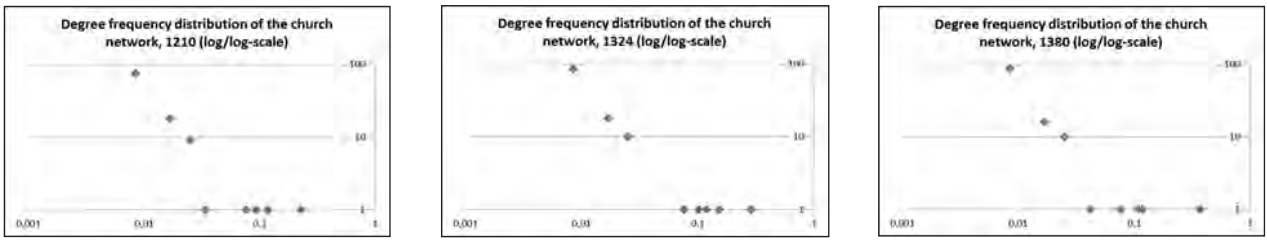

*Figure 15 a-c. The degree frequency distribution of the church administration network in time, 1210 – 1324 – 1380 (on double-logarithmic scales).*

|  | Overlap 1324 | Overlap 1210 | Overlap 1380 |
|---|---|---|---|
| Admin/ Church | 0.1568 | 0.133 | 0.0944 |
| Admin/ Street | 0.0856 | 0.0856 | 0.0859 |
| Church/ Street | 0.1497 | 0.1449 | 0.1438 |
| All 3 | 0.0292 | 0.0229 | 0.027 |

*Figure 16. Overlap between the street, state and church networks for Thrace, 1324 -1210 – 1380.*

frequency distributions of the administrative networks, which again indicate their similar underlying architecture (Fig. 14 a-c, Fig. 15 a-c). In addition, we calculate a high degree and betweenness correlation between the administrative networks and a relatively high network overlap for 1210, when political and ecclesiastical spheres of influence converged (Fig. 16 and 17 a-b); in contrast, for 1380, these coefficients of correlation and the network **overlap between state and church network are significantly lower than in 1324, since political and ecclesiastical borders were not congruent any more** (Fig. 16 and 17 a-b). All in all, we observe a high continuity of underlying architectures of our multiplex network of traffic and administrative infrastructure during the entire period, while the political upheavals of these centuries have their effect on the actual distribution of ties and central network positions and the correlation between state and ecclesiastical administrative networks in these respects.

## Networks as Frameworks for Social Interaction

All three networks analysed by us – streets, state and ecclesiastical administration – constitute frameworks for actual human interaction; traders use existing roads, subordinates receive instructions from their superiors within the administrative hierarchy, etc. Yet not all social interactions will take place along the lines defined in our documents and modelled in our networks, and actual ties between individuals may modify the character of a specific infrastructure network. One example from the ecclesiastical sphere may illustrate this: while the organizational framework of the Byzantine Church was a highly hierarchical one, actual decision-making took place in the

|  | **Degree correlation (r/r²)** 1324 | **Degree correlation (r/r²)** 1210 | **Degree correlation (r/r²)** 1380 |
|---|---|---|---|
| **Admin/Church** | 0.8528/0.7273 | 0.8484/0.7198 | 0.6014/0.3617 |
| **Admin/Street** | 0.2384/0.0568 | 0.2370/0.0562 | 0.3434/0.1179 |
| **Church/Street** | 0.3042/0.0925 | 0.2048/0.0419 | 0.2642/0.0698 |

|  | **Between. correl. (r/r²)** 1324 | **Between. correl. (r/r²)** 1210 | **Between. correl. (r/r²)** 1380 |
|---|---|---|---|
| **Admin/Church** | 0.9108/0.8295 | 0.8425/0.7098 | 0.4248/0.1805 |
| **Admin/Street** | 0.0678/0.0046 | 0.0709/0.0050 | 0.2934/0.0861 |
| **Church/Street** | 0.0825/0.0068 | (-)0.0161/2.6035 | 0.0608/0.0037 |

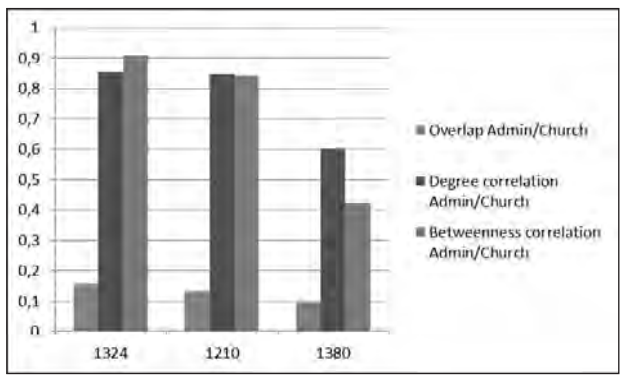

*Figure 17 a-b. Degree and betweenness correlations between the street, state and church networks for Thrace, 1324 – 1210 – 1380.*

synod of Constantinople, where hierarchs from all over the Patriarchate met together with the Patriarch; their directives then where executed through the hierarchical network. In the so-called "Register of the Patriarchate of Constantinople", we find hundreds of such decisions for the period between 1315 and 1402, which also include lists of participants of the respective synodal sessions (Preiser-Kapeller 2008b). We chose one of these sessions from the year 1380, modelled a network of the participating hierarchs and connected it with our ecclesiastical network for the same year (Fig. 18) (cf. also Preiser-Kapeller forthcoming).

We detect now the cluster of the synod in the centre of the church network; some of the participating hierarchs came from bishoprics in Thrace, which are therefore also part of this cluster. This modification of course has also effects on network measures: the density and the clustering coefficient of the network increase (Fig. 19), and the synodal cluster can also be identified in the degree distribution (Fig. 20). Although the effect is primarily a local one – density and clustering increase at the centre, while the rest of the hierarchical network remains unchanged – the impact is significant.

**A Combination of Networks as a "Small World" Model**

Finally, we wanted to analyse the accumulative topology and characteristics of our multiplex infrastructure network; for this purpose, we combined the three networks (street, state, church) for 1324 in one network with one category of links (Fig. 21).

We receive a network which combines characteristics of the original networks in a novel way. The degree frequency distribution shows a relatively equal pattern in the lower values (as does the street network), but a long tail to the right (as the administrative networks) (Fig.22). Network centralization is high, average path length is low (as in the administrative networks); but the combined density and especially the clustering coefficient are significantly higher than in any of the three networks (Fig. 23). All in all, the combined network shows the characteristics of a "small world-network", as Watts and Strogatz have established it (Watts 1999; Watts and Strogatz 1998, 440-442); it has a relatively low average path length (meaning higher connectivity within the network) and a relatively high clustering coefficient (meaning high local density of network ties), uniting effectiveness and robustness. Thus, the combined infrastructural network follows a well-established model for real-world networks, which has been analysed for modern day and past periods; only the application of the multiplexity-approach allowed us to identify this highly interesting pattern.

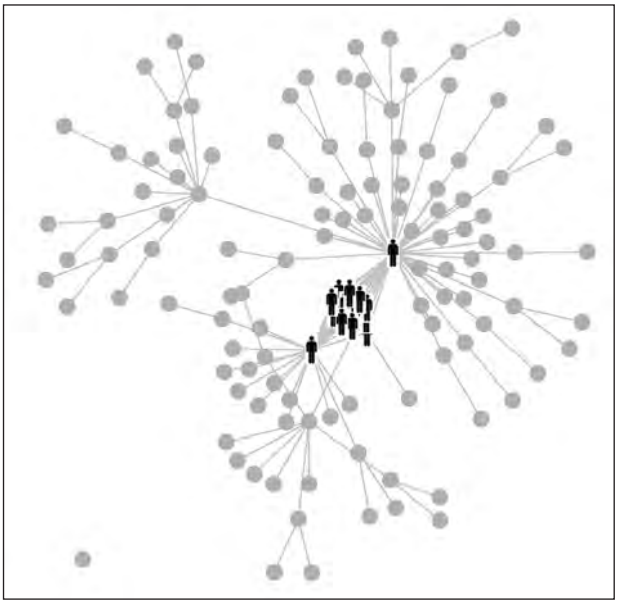

*Figure 18. The combination of the network of the synodal session of June 1380 with the church administrative network for 1380.*

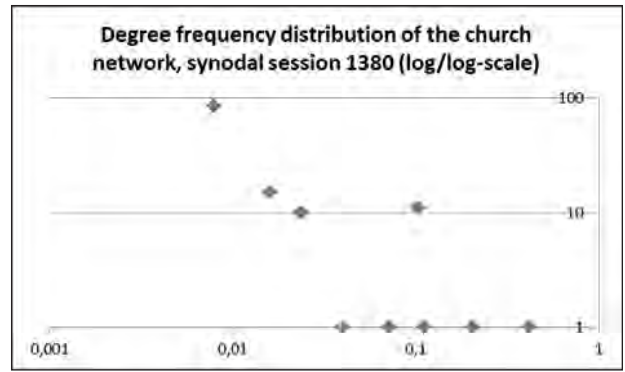

*Figure 20. Degree frequency distribution of the combined network for the synodal session of June 1380.*

**Conclusions**

To sum up: if source evidence allows us to construct multiplex networks on the same set of nodes, tools of modern network analysis and statistics enable us firstly to inspect the respective networks individually; thus, we may find that the different spheres which we define (such as state or religion in our example) span networks with very similar or very different architectures – and that different rules, intentions and purposes of interaction may significantly influence network topologies (which *vice versa* can determine acceptable lines of interaction). Secondly, we can also analyse the correlations and congruence between different networks (and the spheres they represent), especially if we also have the opportunity to observe the same networks for different timeframes. And thirdly, we can observe the combined effects of a multiplex network on the topology of social interaction lines - and the emerging framework may be more than only the combination of its parts. The multiplexity approach illustrates the actual complexity of human networks. At the same time, it makes us aware that even if we are able to construct a network for one sphere of human interaction for a specific segment of a society, we cannot assume that the structural properties of individual nodes, cliques or the entire network emerge only due to the topology of the observed network, but we have to account for the correlations within the multiplexity of relational spheres which we are not able to observe; and this is even more true, as Tom Brughmans has stated, if we encounter a "*fragmentary nature of our sources*" and are ignorant "*of the entire population our sample is derived from*" (Brughmans forthcoming). Modern network analysis is not a tool of reductionist structuralism, but a window to the complexity and diversity of human society, which challenges the imaginativeness of archaeologists and historians.

| Measure | Church 1210 | Church 1324 | Church 1380 | Synod 1380 |
|---|---|---|---|---|
| Average Distance | 3.011 | 3.532 | 3.375 | 3.25 |
| Clustering Coefficient, Watts-Strogatz | 0 | 0 | 0 | 0.095 |
| Density | 0.014 | 0.017 | 0.017 | 0.025 |
| Link Count | 202 | 234 | 234 | 408 |
| Network Centralization, Betweenness | 0.192 | 0.865 | 0.899 | 0.874 |
| Network Centralization, Total Degree | 0.218 | 0.276 | 0.362 | 0.398 |

*Figure 19. Comparison of the church networks for 1210, 1324 and 1380 with the combined network for the synodal session of June 1380.*

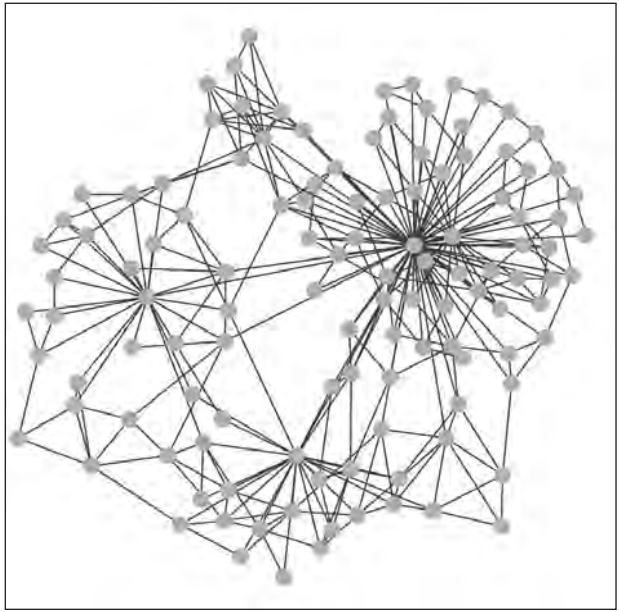

*Figure 21. The combined "infrastructure network" for Thrace for 1324.*

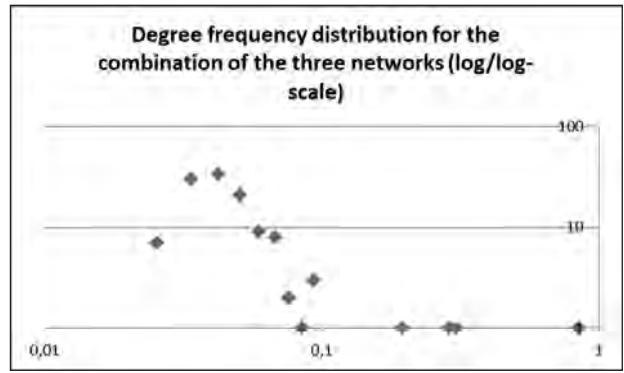

*Figure 22. Degree frequency distribution for the combination of the three networks for 1324 (on double-logarithmic scale).*

| Measure | Admin 1324 | Church 1324 | Street | Combination 1324 |
|---|---|---|---|---|
| Average Distance | 2.779 | 3.532 | 7.085 | 3.458 |
| Clustering Coefficient, Watts-Strogatz | 0 | 0 | 0.092 | 0.55 |
| Density | 0.017 | 0.017 | 0.025 | 0.045 |
| Link Count | 234 | 234 | 350 | 632 |
| Network Centralization, Betweenness | 0.836 | 0.865 | 0.196 | 0.701 |
| Network Centralization, Total Degree | 0.509 | 0.276 | 0.035 | 0.803 |

*Figure 23. A comparison of network analytical measures for the state, church and street network of 1324 with their combination.*

## Acknowledgements

This study was undertaken as part of the Project "The term of office of Patriarch Antonios IV (1391-1397) of Constantinople as reflected in the documents of the Registrum Patriarchatus Constantinopolitani", Austrian Science Fund (FWF-Project P22269); project director is Prof. Otto Kresten (Vienna). Many thanks also to Michele Coscia (Pisa and Boston), who generously sent a draft of his study on the "Foundations of Multidimensional Network Analysis" to me. Networks were created with the software packages *Pajek*\* (URL: http://vlado.fmf.uni-lj.si/pub/networks/pajek/) and *ORA*\* (URL: http://www.casos.cs.cmu.edu/projects/ora/).

# Colour versions of figures

**Figures**

**Fig 1:** Topological representation of the network of roads in the area of Thrace in the 13[th]/14[th] cent. (119 nodes or localities and 350 links or routes)

**Fig. 2:** Topological representation of the searoute-network for the area of Thrace in the 13[th]/14[th] cent. (32 nodes or localities and 70 links or routes)

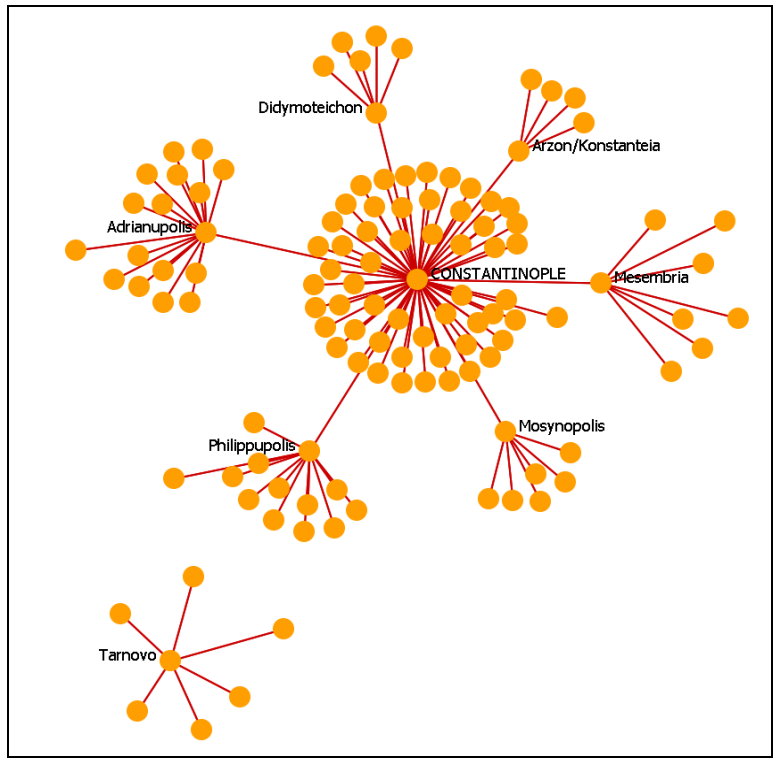

**Fig. 3:** Topological representation of the network of administration of the states for the area of Thrace in 1324 (119 nodes and 234 links between administrative centres and their subordinate localities)

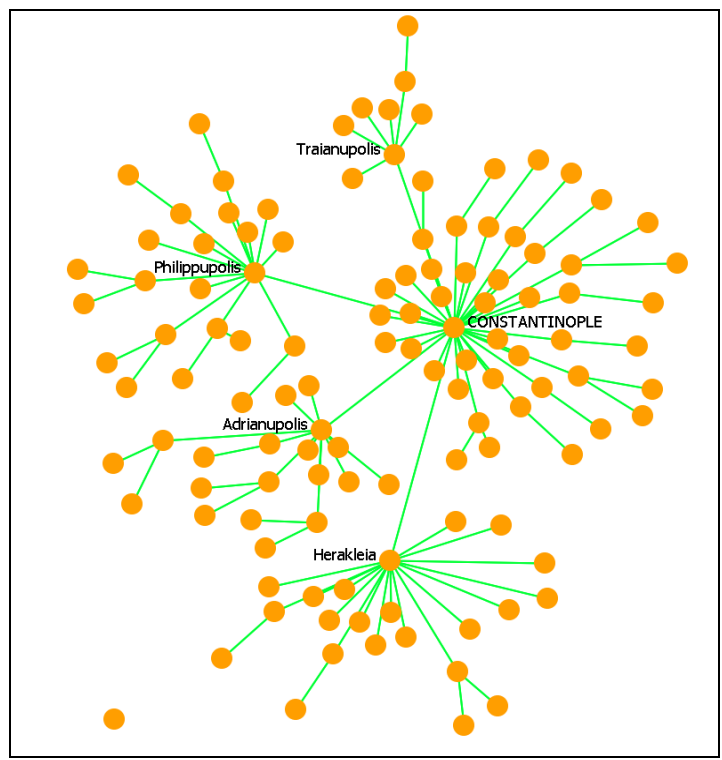

**Fig. 4:** Topological representation of the network of administration of the churches for the area of Thrace in 1324 (119 node and 234 links between the Patriarchate, metropolitan sees, archbishoprics and suffragan bishoprics)

| Measure | Admin 1324 | Church 1324 | Street |
|---|---|---|---|
| **Average Distance** | 2.779 | 3.532 | 7.085 |
| **Clustering Coefficient, Watts-Strogatz** | 0 | 0 | 0,092 |
| **Density** | 0.017 | 0.017 | 0.025 |
| **Link Count** | 234 | 234 | 350 |
| **Network Centralization, Betweenness** | 0.836 | 0.865 | 0.196 |
| **Network Centralization, Total Degree** | 0.509 | 0.276 | 0.035 |

**Fig. 5:** A comparison of network analytical measures for the road, state and church networks for Thrace for 1324.

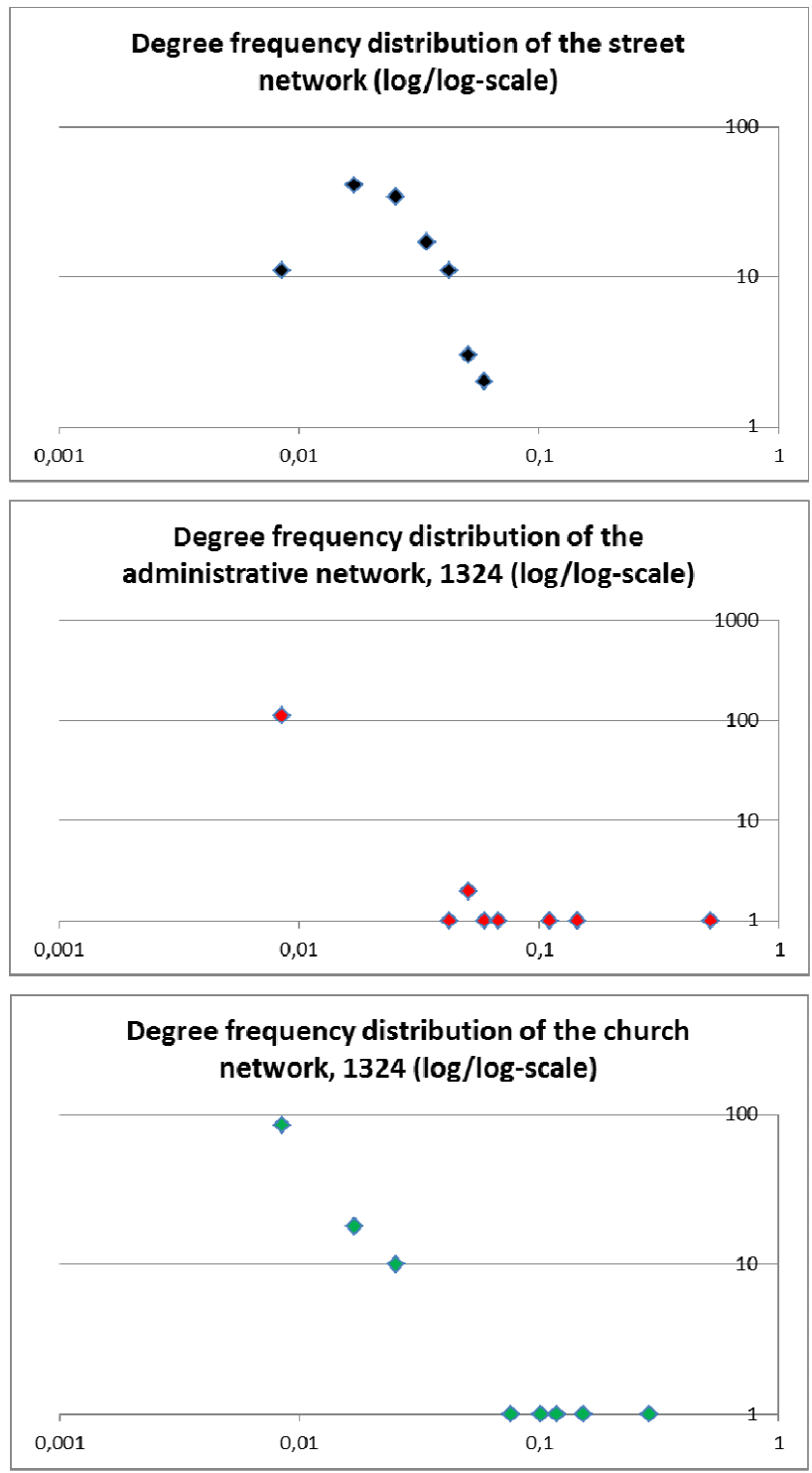

**Fig. 6 a-c:** A comparison of the road, state and church networks for Thrace for 1324: degree frequency distributions (on double-logarithmic scales; we measure the frequency of each degree values; on "degree", see section 3)

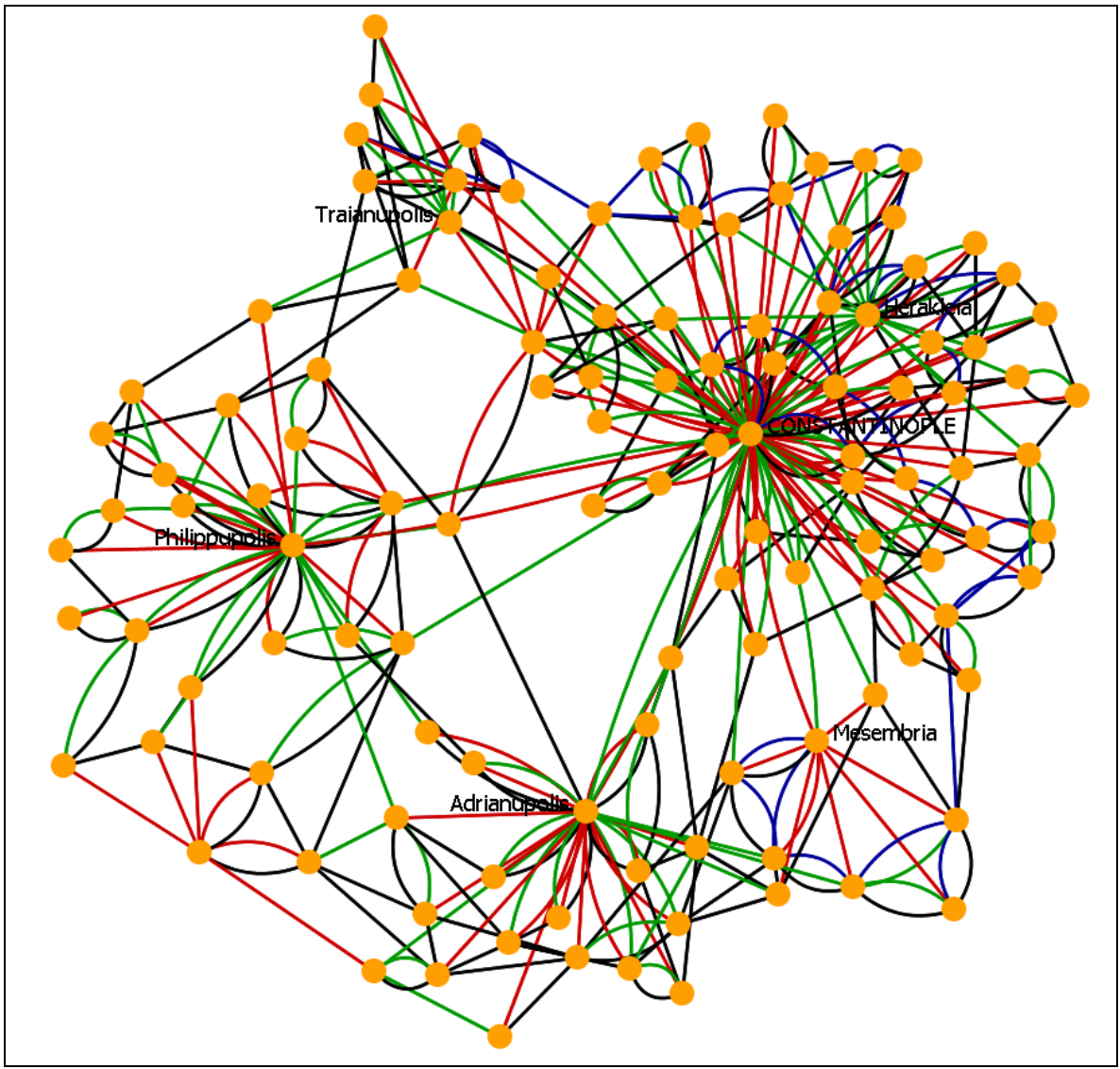

**Fig. 7:** The multiplex „infrastructure" network of Thrace, 1324 (**black: street network**; **blue: searoutes-network**; **red: state administration network**; **green: church administration network**)

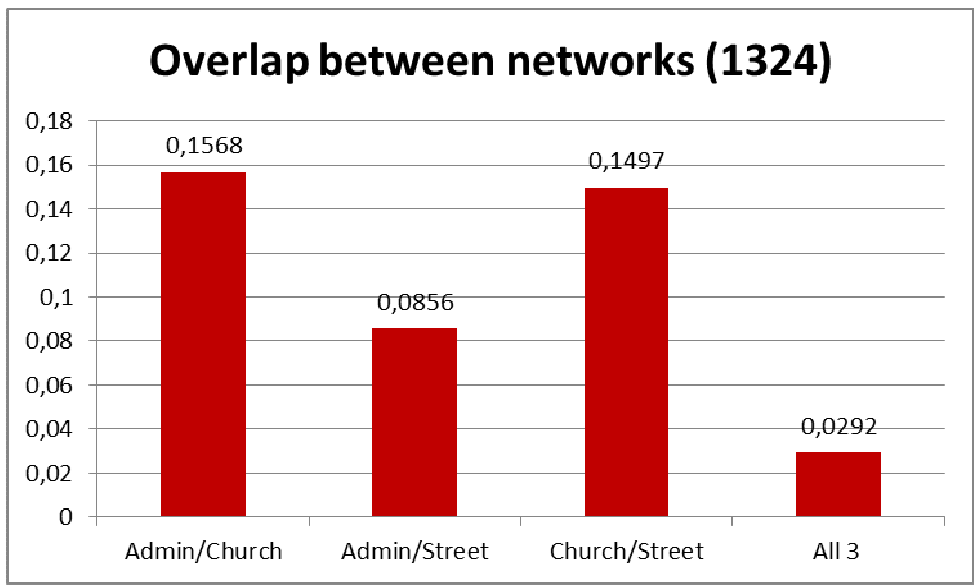

**Fig. 8:** The overlap (percentage of links between nodes common to two or all three networks) between the networks for Thrace in 1324

|  | Degree correlation (r/r²) 1324 | Between. correl. (r/r²) 1324 |
|---|---|---|
| **Admin/Church** | 0.8528/0.7273 | 0.9108/0.8295 |
| **Admin/Street** | 0.2384/0.0568 | 0.0678/0.0046 |
| **Church/Street** | 0.3042/0.0925 | 0.0825/0.0068 |

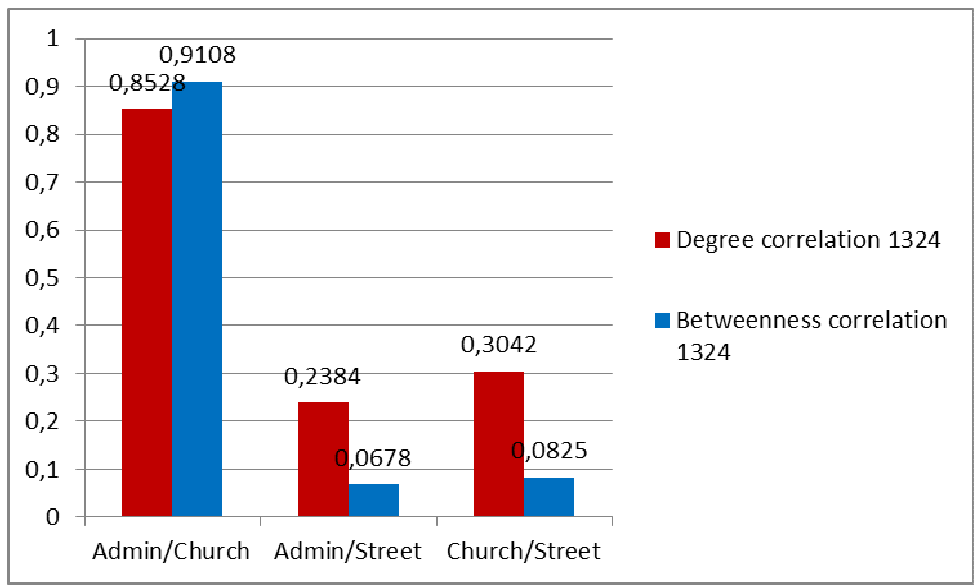

**Fig. 9 a and b:** Degree and betweenness correlations between the networks for Thrace in 1324; on degree and betweenness, see section 3.

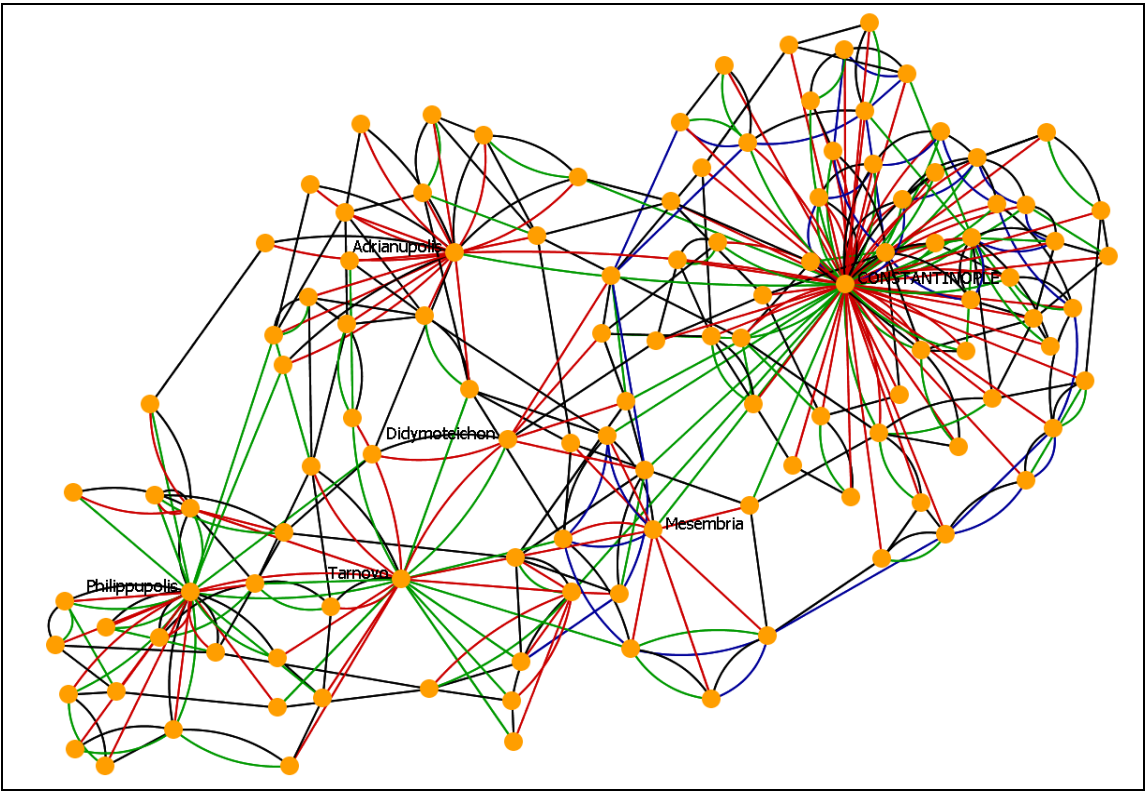

**Fig. 10:** The multiplex „infrastructure" network of Thrace, 1210 (**black: street network**; **blue: searoutes-network**; **red: state administration network**; **green: church administration network**)

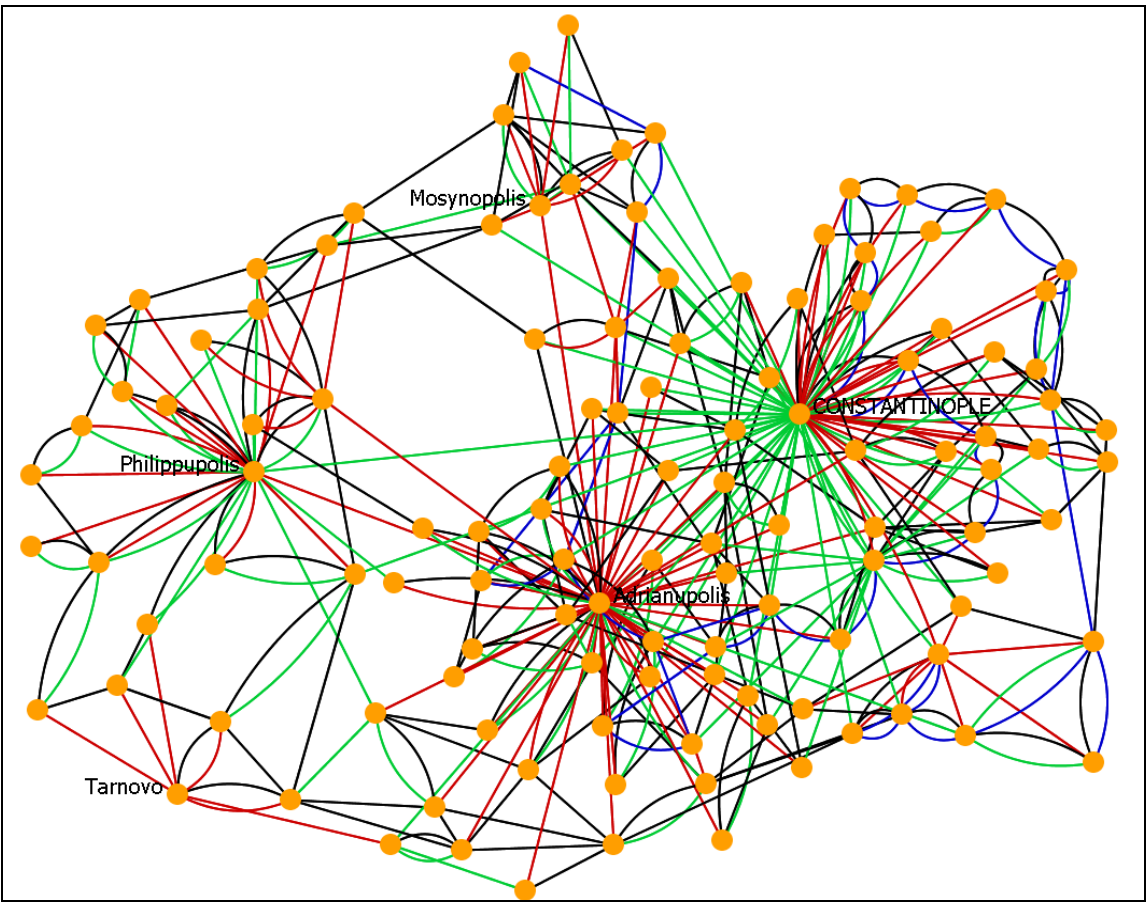

**Fig. 11:** The multiplex „infrastructure" network of Thrace, 1380 (**black: street network**; **blue: searoutes-network**; **red: state administration network**; **green: church administration network**)

| Measure | Admin 1210 | Admin 1324 | Admin 1380 |
|---|---|---|---|
| Average Distance | 2.543 | 2.779 | 2.528 |
| Clustering Coefficient, Watts-Strogatz | 0 | 0 | 0 |
| Density | 0.017 | 0.017 | 0.017 |
| Link Count | 234 | 234 | 232 |
| Network Centralization, Betweenness | 0.347 | 0.836 | 0.381 |
| Network Centralization, Total Degree | 0.466 | 0.509 | 0.397 |

| Measure | Church 1210 | Church 1324 | Church 1380 |
|---|---|---|---|
| Average Distance | 3.011 | 3.532 | 3.375 |
| Clustering Coefficient. Watts-Strogatz | 0 | 0 | 0 |
| Density | 0.014 | 0.017 | 0.017 |
| Link Count | 202 | 234 | 234 |
| Network Centralization. Betweenness | 0.192 | 0.865 | 0.899 |
| Network Centralization. Total Degree | 0.218 | 0.276 | 0.362 |

**Fig. 12:** A comparison of network analytical measures for the administrative networks in time, 1210 – 1324 – 1380

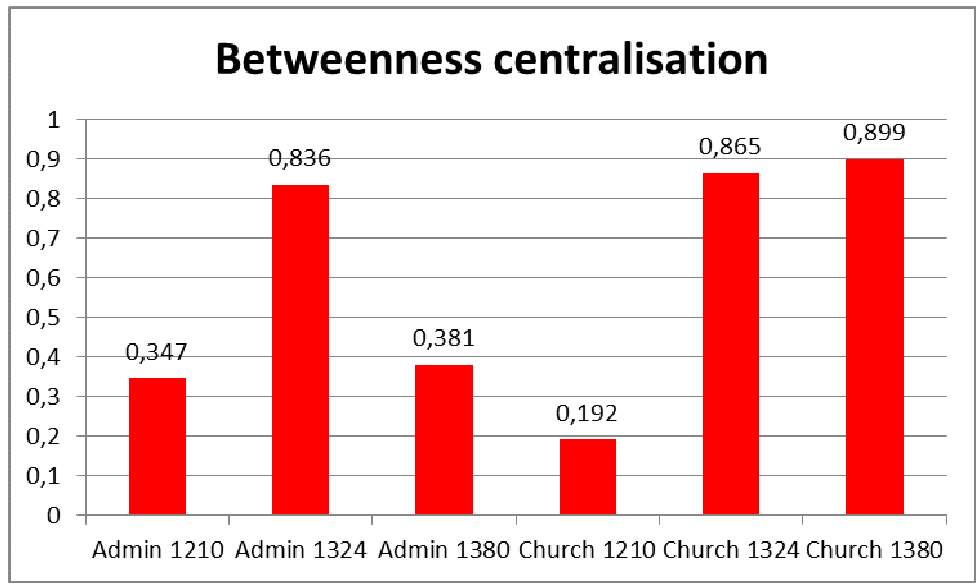

**Fig. 13:** A comparison of betweenness centralization for the administrative networks in time, 1210 – 1324 – 1380

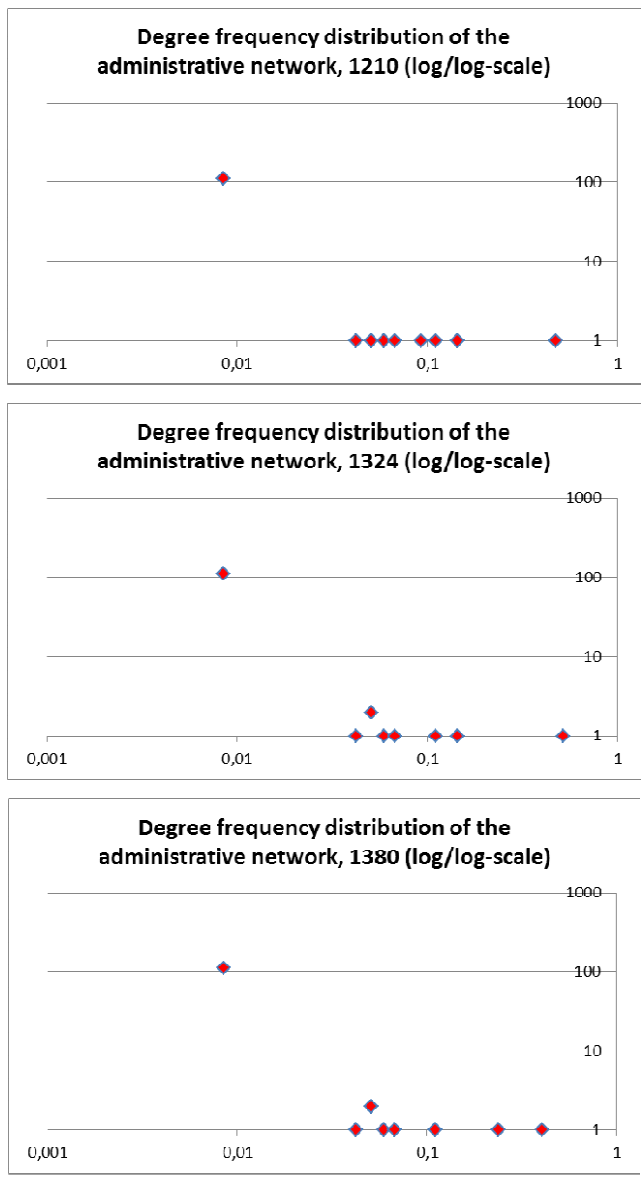

**Fig. 14 a-c:** The degree frequency distribution of the state administration network in time, 1210 – 1324 – 1380 (on double-logarithmic scales)

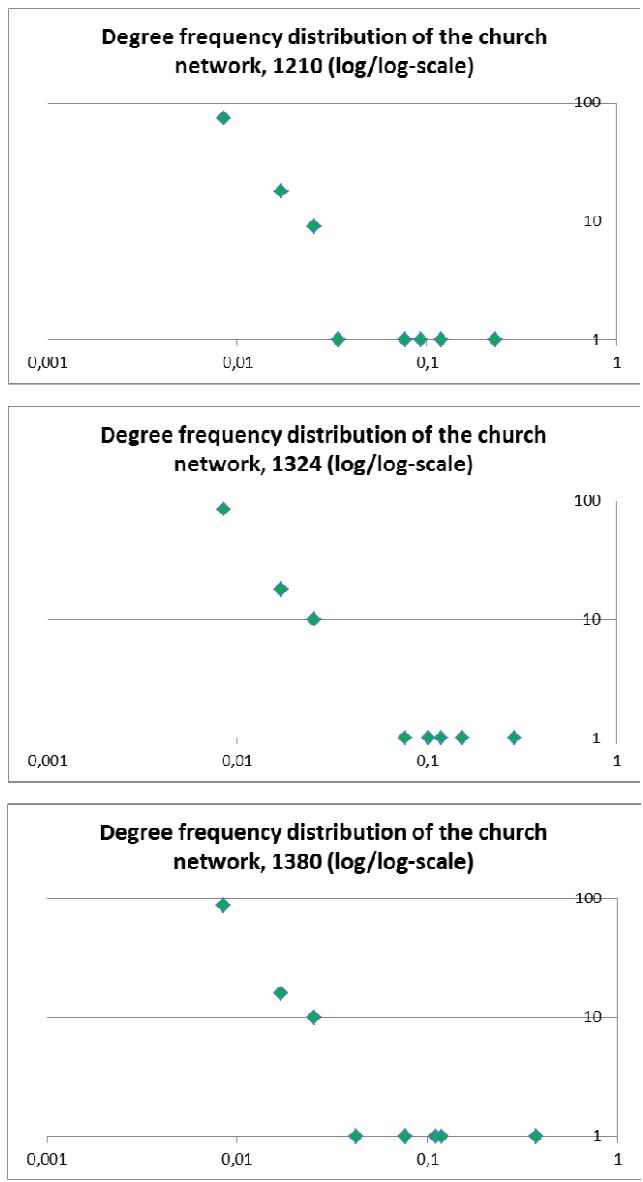

**Fig. 15 a-c:** The degree frequency distribution of the church administration network in time, 1210 – 1324 – 1380 (on double-logarithmic scales)

|  | Overlap 1324 | Overlap 1210 | Overlap 1380 |
|---|---|---|---|
| **Admin/Church** | 0.1568 | 0.133 | 0.0944 |
| **Admin/Street** | 0.0856 | 0.0856 | 0.0859 |
| **Church/Street** | 0.1497 | 0.1449 | 0.1438 |
| **All 3** | 0.0292 | 0.0229 | 0.027 |

**Fig. 16:** Overlap between the street, state and church networks for Thrace, 1324 – 1210 – 1380

|                | Degree correlation (r/r²) 1324 | Degree correlation (r/r²) 1210 | Degree correlation (r/r²) 1380 |
|----------------|-------------------------------|-------------------------------|-------------------------------|
| **Admin/Church**   | 0.8528/0.7273 | 0.8484/0.7198 | 0.6014/0.3617 |
| **Admin/Street**   | 0.2384/0.0568 | 0.2370/0.0562 | 0.3434/0.1179 |
| **Church/Street**  | 0.3042/0.0925 | 0.2048/0.0419 | 0.2642/0.0698 |

|                | Between. correl. (r/r²) 1324 | Between. correl. (r/r²) 1210 | Between. correl. (r/r²) 1380 |
|----------------|-----------------------------|-----------------------------|-----------------------------|
| **Admin/Church**   | 0.9108/0.8295 | 0.8425/0.7098 | 0.4248/0.1805 |
| **Admin/Street**   | 0.0678/0.0046 | 0.0709/0.0050 | 0.2934/0.0861 |
| **Church/Street**  | 0.0825/0.0068 | (-)0.0161/2.6035 | 0.0608/0.0037 |

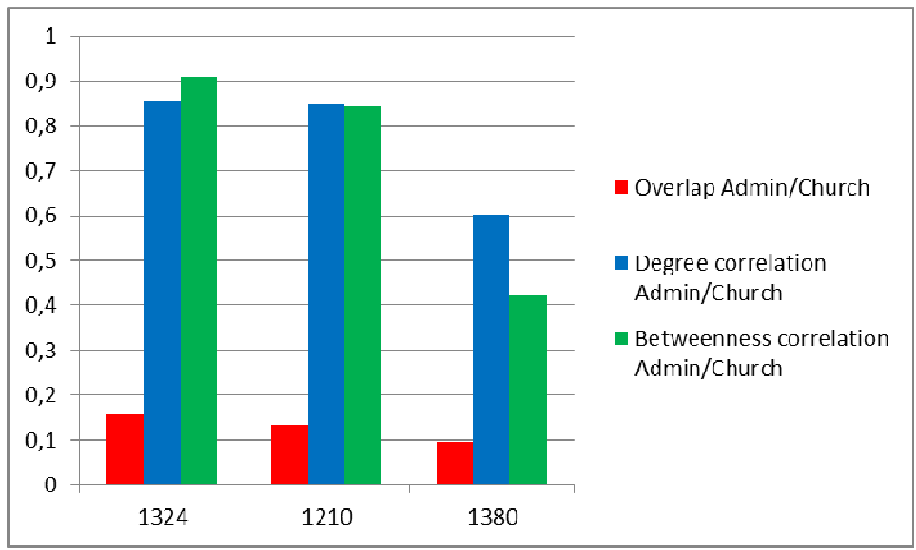

**Fig. 17 a-b:** Degree and betweenness correlations between the street, state and church networks for Thrace, 1324 – 1210 – 1380

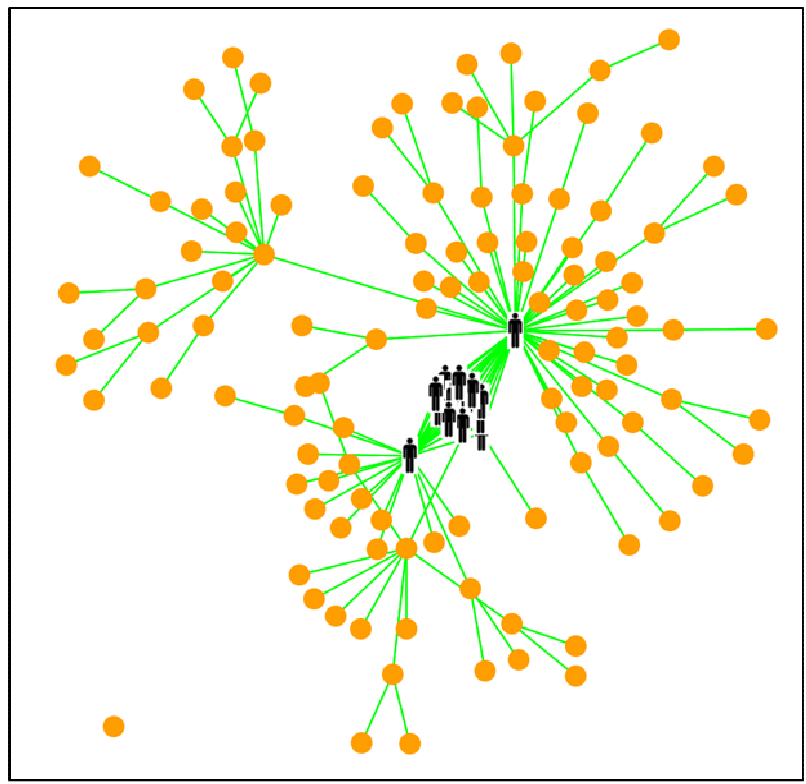

**Fig. 18:** The combination of the network of the synodal session of June 1380 with the church administrative network for 1380

| Measure | Church 1210 | Church 1324 | Church 1380 | Synod 1380 |
|---|---|---|---|---|
| Average Distance | 3.011 | 3.532 | 3.375 | 3.25 |
| Clustering Coefficient, Watts-Strogatz | 0 | 0 | 0 | 0.095 |
| Density | 0.014 | 0.017 | 0.017 | 0.025 |
| Link Count | 202 | 234 | 234 | 408 |
| Network Centralization, Betweenness | 0.192 | 0.865 | 0.899 | 0.874 |
| Network Centralization, Total Degree | 0.218 | 0.276 | 0.362 | 0.398 |

**Fig. 19:** Comparison of the church networks for 1210, 1324 and 1380 with the combined network for the synodal session of June 1380

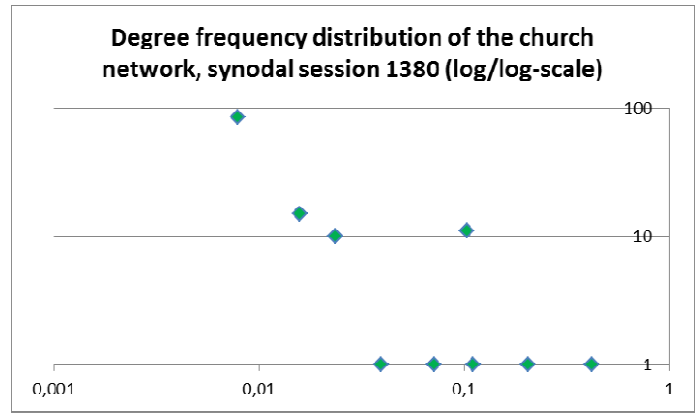

**Fig. 20:** Degree frequency distribution of the combined network for the synodal session of June 1380

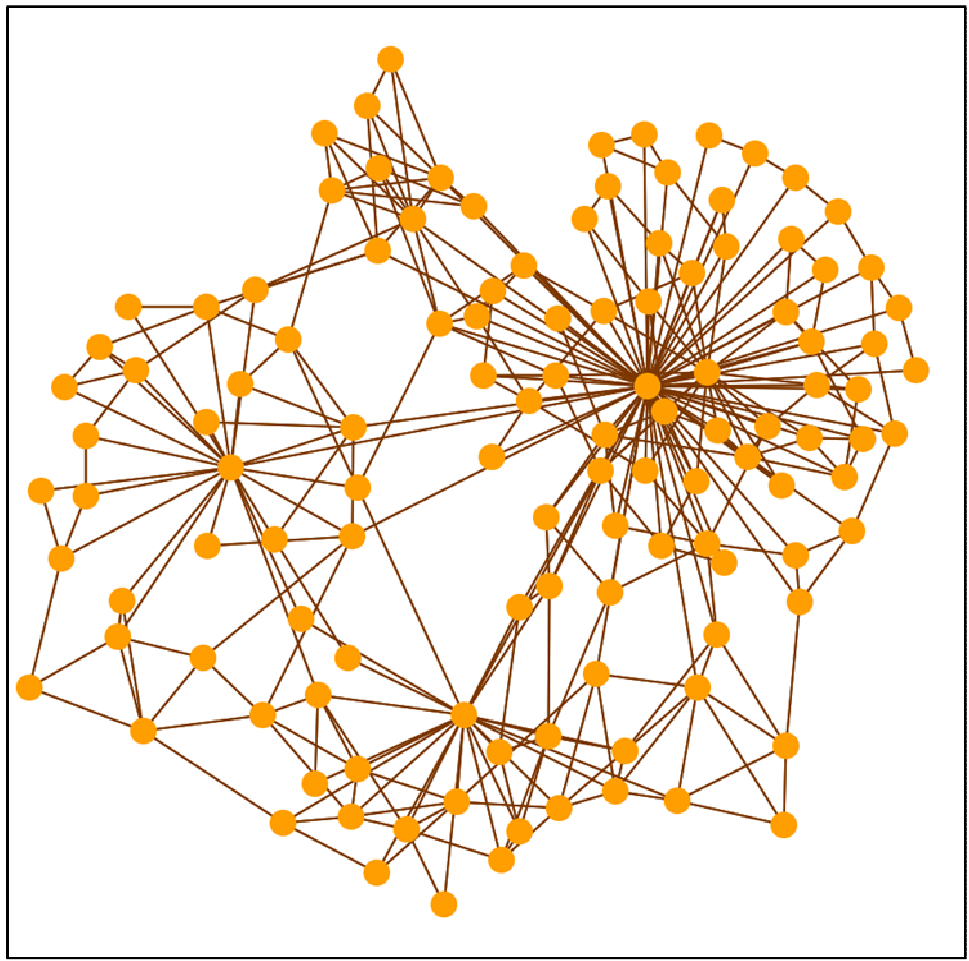

**Fig. 21:** The combined "infrastructure network" for Thrace for 1324

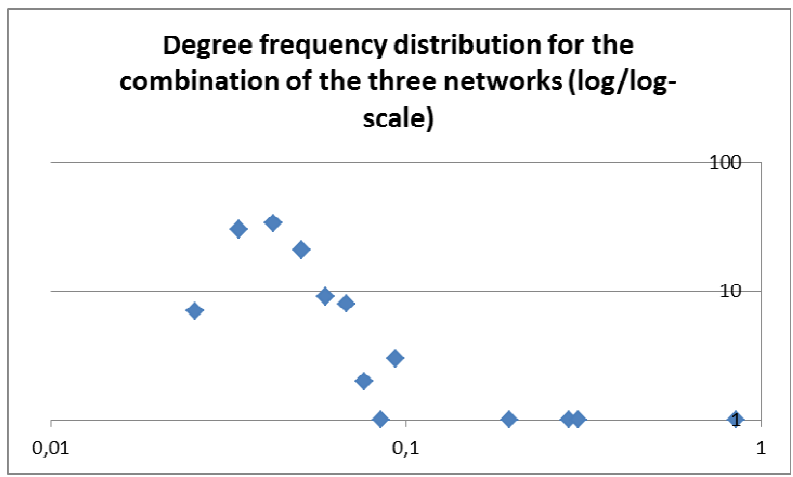

**Fig. 22:** Degree frequency distribution for the combination of the three networks for 1324 (on double-logarithmic scale)

| Measure | Admin 1324 | Church 1324 | Street | Combination 1324 |
|---|---|---|---|---|
| Average Distance | 2.779 | 3.532 | 7.085 | 3.458 |
| Clustering Coefficient, Watts-Strogatz | 0 | 0 | 0.092 | 0.55 |
| Density | 0.017 | 0.017 | 0.025 | 0.045 |
| Link Count | 234 | 234 | 350 | 632 |
| Network Centralization, Betweenness | 0.836 | 0.865 | 0.196 | 0.701 |
| Network Centralization, Total Degree | 0.509 | 0.276 | 0.035 | 0.803 |

**Fig. 23:** A comparison of network analytical measures for the state, church and street network of 1324 with their combination